\documentclass[onecolumn,showpacs,preprintnumbers,amsmath,amssymb,eqsecnum]{revtex4}
\usepackage{dcolumn}
\usepackage{bm}
\usepackage[dvips]{graphicx}
\usepackage{wrapfig}
\usepackage{psfrag}
\usepackage{color}
\begin{document}


\def\sh{\mathop{\rm sh}\nolimits}
\def\ch{\mathop{\rm ch}\nolimits}
\def\var{\mathop{\rm var}}\def\exp{\mathop{\rm exp}\nolimits}
\def\Re{\mathop{\rm Re}\nolimits}
\def\Sp{\mathop{\rm Sp}\nolimits}
\def\kp{\mathop{\text{\ae}}\nolimits}
\def\bk{{\bf {k}}}
\def\bp{{\bf {p}}}
\def\bq{{\bf {q}}}
\def\lra{\mathop{\longrightarrow}}
\def\Const{\mathop{\rm Const}\nolimits}
\def\sh{\mathop{\rm sh}\nolimits}
\def\ch{\mathop{\rm ch}\nolimits}
\def\var{\mathop{\rm var}}
\def\mK{\mathop{{\mathfrak {K}}}\nolimits}
\def\mR{\mathop{{\mathfrak {R}}}\nolimits}
\def\mv{\mathop{{\mathfrak {v}}}\nolimits}
\def\mV{\mathop{{\mathfrak {V}}}\nolimits}
\def\mD{\mathop{{\mathfrak {D}}}\nolimits}
\def\mN{\mathop{{\mathfrak {N}}}\nolimits}
\def\mS{\mathop{{\mathfrak {S}}}\nolimits}
\def\mP{\mathop{{\mathfrak {P}}}\nolimits}
\def\mL{\mathop{{\mathfrak {L}}}\nolimits}
\def\mM{\mathop{{\mathfrak {M}}}\nolimits}

\newcommand\ve[1]{{\mathbf{#1}}}

\def\Re{\mbox {Re}}
\newcommand{\Z}{\mathbb{Z}}
\newcommand{\R}{\mathbb{R}}
\def\mK{\mathop{{\mathfrak {K}}}\nolimits}
\def\mk{\mathop{{\mathfrak {k}}}\nolimits}
\def\mR{\mathop{{\mathfrak {R}}}\nolimits}
\def\mv{\mathop{{\mathfrak {v}}}\nolimits}
\def\mV{\mathop{{\mathfrak {V}}}\nolimits}
\def\mD{\mathop{{\mathfrak {D}}}\nolimits}
\def\mN{\mathop{{\mathfrak {N}}}\nolimits}
\def\ml{\mathop{{\mathfrak {l}}}\nolimits}
\def\mf{\mathop{{\mathfrak {f}}}\nolimits}
\def\mF{\mathop{{\mathfrak {F}}}\nolimits}
\newcommand{\ccm}{{\cal M}}
\newcommand{\cE}{{\cal E}}
\newcommand{\cV}{{\cal V}}
\newcommand{\cI}{{\cal I}}
\newcommand{\cR}{{\cal R}}
\newcommand{\cK}{{\cal K}}
\newcommand{\cH}{{\cal H}}
\newcommand{\cW}{{\cal W}}
\newcommand{\cL}{{\cal L}}
\newcommand{\cF}{{\cal F}}

\def\br{\mathop{{\bf {r}}}\nolimits}
\def\bS{\mathop{{\bf {S}}}\nolimits}
\def\bA{\mathop{{\bf {A}}}\nolimits}
\def\bJ{\mathop{{\bf {J}}}\nolimits}
\def\bn{\mathop{{\bf {n}}}\nolimits}
\def\bg{\mathop{{\bf {g}}}\nolimits}
\def\bv{\mathop{{\bf {v}}}\nolimits}
\def\be{\mathop{{\bf {e}}}\nolimits}
\def\bp{\mathop{{\bf {p}}}\nolimits}
\def\bz{\mathop{{\bf {z}}}\nolimits}
\def\bbf{\mathop{{\bf {f}}}\nolimits}
\def\bb{\mathop{{\bf {b}}}\nolimits}
\def\ba{\mathop{{\bf {a}}}\nolimits}
\def\bx{\mathop{{\bf {x}}}\nolimits}
\def\by{\mathop{{\bf {y}}}\nolimits}
\def\br{\mathop{{\bf {r}}}\nolimits}
\def\bs{\mathop{{\bf {s}}}\nolimits}
\def\bH{\mathop{{\bf {H}}}\nolimits}
\def\bk{\mathop{{\bf {k}}}\nolimits}
\def\be{\mathop{{\bf {e}}}\nolimits}
\def\bnul{\mathop{{\bf {0}}}\nolimits}
\def\bq{{\bf {q}}}

\newcommand{\oV}{\overline{V}}
\newcommand{\vkp}{\varkappa}
\newcommand{\os}{\overline{s}}
\newcommand{\opsi}{\overline{\psi}}
\newcommand{\ov}{\overline{v}}
\newcommand{\oW}{\overline{W}}
\newcommand{\oPhi}{\overline{\Phi}}

\def\mI{\mathop{{\mathfrak {I}}}\nolimits}
\def\mA{\mathop{{\mathfrak {A}}}\nolimits}

\def\st{\mathop{\rm st}\nolimits}
\def\tr{\mathop{\rm tr}\nolimits}
\def\sign{\mathop{\rm sign}\nolimits}
\def\d{\mathop{\rm d}\nolimits}
\def\const{\mathop{\rm const}\nolimits}
\def\diag{\mathop{\rm diag}\nolimits}
\def\O{\mathop{\rm O}\nolimits}
\def\Spin{\mathop{\rm Spin}\nolimits}
\def\exp{\mathop{\rm exp}\nolimits}
\def\SU{\mathop{\rm SU}\nolimits}
\def\mU{\mathop{{\mathfrak {U}}}\nolimits}
\newcommand{\cU}{{\cal U}}
\newcommand{\cD}{{\cal D}}

\def\mI{\mathop{{\mathfrak {I}}}\nolimits}
\def\mA{\mathop{{\mathfrak {A}}}\nolimits}
\def\mU{\mathop{{\mathfrak {U}}}\nolimits}

\def\st{\mathop{\rm st}\nolimits}
\def\tr{\mathop{\rm tr}\nolimits}
\def\sign{\mathop{\rm sign}\nolimits}
\def\d{\mathop{\rm d}\nolimits}
\def\const{\mathop{\rm const}\nolimits}
\def\O{\mathop{\rm O}\nolimits}
\def\Spin{\mathop{\rm Spin}\nolimits}
\def\exp{\mathop{\rm exp}\nolimits}

\title{Phase transition near the Big Bang in the lattice theory of gravity}

\author {S.N. Vergeles\vspace*{4mm}\footnote{{e-mail:vergeles@itp.ac.ru}}}

\affiliation{Landau Institute for Theoretical Physics,
Russian Academy of Sciences,
Chernogolovka, Moscow region, 142432 Russia \linebreak
and   \linebreak
Moscow Institute of Physics and Technology, Department
of Theoretical Physics, Dolgoprudnyj, Moskow region,
141707 Russia}

\begin{abstract}
Lattice regularization of the theory of gravity provides a new possibility for study of the Big Bang problem. In the 4D lattice theories of gravity, the existence of a high-temperature phase is proved, which is characterized by the folding of space into a point and zeroing of the energy-momentum tensor. 
The existence of a low-temperature phase in the long-wave limit is also shown, the geometric properties of which and its dynamics correspond to known concepts: the expansion of the Universe first follows an exponential law, and then smoothly transitions to a power-law regime.
\end{abstract}

\pacs{04.60.Bc}

\maketitle

\section{Introduction}

In the works of the author \cite{vergeles2006one,vergeles2015wilson,vergeles2017fermion,vergeles2017note,vergeles2021note} a model of discrete gravity on a simplicial complex was studied. This model included pure gravity degrees of freedom (tetrads, connection), as well as a fermion Dirac field minimally coupled to gravity. In particular, in the work \cite{vergeles2021note} a discrete symmetry $\Z_2$ of the action called PT symmetry was described. This PT transformation changes the sign of all tetrads and mutually interchanges the Dirac variables with their Hermitian conjugates, that is, mutually interchanges particles and antiparticles.

In this work, we take the next step: we show that at ultrahigh temperatures the PT symmetry is not broken, but it is broken at sufficiently low temperatures. In the high-temperature phase, the mean values of the energy-momentum tensor of matter, as well as the mean values of the tetrads, are equal to zero. On the contrary, in the low-temperature phase these values are different from zero. Tetrad is an order parameter.

As is known, in the Minimal Standard Model (MSM) the number of fermionic degrees of freedom is multiple than the number of bosonic degrees of freedom (even taking into account two gravitons). It follows that the total energy of vacuum zero-point oscillations is negative. We assume that some properties of the MSM
near the point of the Big Bang can be described using the discrete gravity model being studied.
Here we are interested in the following question: can the continuation of solutions to Einstein’s continuum equations in the vicinity of the Big Bang point contain some information about the properties of the high-temperature phase in the lattice theory? To answer this question, we solve Einstein's equations in the spirit of Friedmann's model, but with a bare cosmological constant of positive sign.
The positive cosmological constant compensates for the enormous negative energy of the vacuum.
And since we consider the continuum theory as the long-wave limit of the discrete theory of gravity, then all the quantities in the equations (vacuum energy density $\varepsilon$, the bare cosmological constant $\Lambda_0$, etc.) are finite. In this case, Einstein's equations and their solutions are mathematically correct. Moreover, the solutions obtained demonstrate the necessary properties common to cosmology: near the point of the Big Bang, an exponential expansion regime of the Universe takes place (inflation phase), but then the expansion switches to a power-law regime. It is also interesting that in the solution found, the vacuum energy modulus $|\varepsilon|$ decreases as we approach the Big Bang point. In other words, the energy-momentum tensor of matter in the continuum theory tends to zero when approaching the Big Bang point.
In our opinion, this trend suggests that the Einstein continuum equation under consideration actually models a discrete theory of gravity in the long-wave limit. However, one must keep in mind that as the absolute value of the energy-momentum tensor decreases, the role of quantum fluctuations increases, and the classical Einstein equations become inapplicable (see Section IV).

The article is organized as follows.

Since different variants of simplicial theories of gravity are discussed, for ease of reading we define the variant studied here in Section II.

Section III proves the main result of the work: in the lattice theory of gravity, discrete PT symmetry is not broken at ultrahigh temperatures. But at low temperatures, spontaneous breaking of this symmetry occurs.

Section IV considers the solution of the Einstein equation within the Friedmann paradigm in the long-wave limit. This consideration makes sense, since the lattice theory transforms into the ordinary theory of gravity in the long-wave limit. It is shown that the solution of classical equations is possible only at some distance from the point of the Big Bang. The reason for this is that as one approaches the singularity, quantum fluctuations become significant.

Section V provides some thermodynamic considerations.

This work is of a model nature.

\section{introduction to lattice theory of gravity}

\subsection{Lattice theory definition}

To go further, it is necessary to formulate the variant of the theory of gravity on a lattice, which we will use. For a more detailed description of this theory, see
\cite{vergeles2006one,vergeles2015wilson,vergeles2017fermion,vergeles2017note,vergeles2021note}

This subsection uses the Euclidean signature.
Denote by $\gamma^a$ $4\times4$ Hermitian Dirac matrices, so that
\begin{gather}
\gamma^a\gamma^b+\gamma^b\gamma^a=2\delta^{ab}, \quad \sigma^{ab}\equiv\frac{1}{4}[\gamma^a,\gamma^b], \quad a=1,2,3,4, \quad
\gamma^5\equiv\gamma^1\gamma^2\gamma^3\gamma^4=(\gamma^5)^{\dag}.
\label{Dirac_Algebra}
\end{gather}

Consider the orientable abstract 4-dimensional simplicial complex  $\mK$.
Suppose that any of its 4-simplexes belongs to such a finite (or infinite) sub-complex ${\mK}'\in\mK$  which has a geometric realization in  $\R^4$ topologically equivalent to a disk  without cavities.
The vertices are designated as
$a_{\cV}$, the indices ${\cV}=1,2,\dots,\,{\mN}^{(0)}\rightarrow\infty$ and ${\cW}$ enumerate the vertices
and 4-simplices, correspondingly. It is necessary to use
the local enumeration of the vertices $a_{\cV}$ attached to a given
4-simplex: the all five vertices of a 4-simplex with index ${\cW}$
are enumerated as $a_{{\cV}_{({\cW})i}}$, $i=1,2,3,4,5$. The later notations with extra low index  $({\cW})$
indicate that the corresponding quantities belong to the
4-simplex with index ${\cW}$. Note that a quantity belonging to the 4-simplex $s^4_{\cW}$ can also belong to the neighboring 4-simplex $s^4_{\cW'}$ if these 4-simplices have common simplices of lower dimension.
The Levi-Civita symbol with in pairs
different indexes
$\varepsilon_{{\cV}_{({\cW})1}{\cV}_{({\cW})2}{\cV}_{({\cW})3}{\cV}_{({\cW})4}{\cV}_{({\cW})5}}=\pm 1$ depending on
whether the order of vertices
$s^4_{\cW}=a_{{\cV}_{({\cW})1}}a_{{\cV}_{({\cW})2}}a_{{\cV}_{({\cW})3}}a_{{\cV}_{({\cW})4}}a_{{\cV}_{({\cW})5}}$ defines the
positive or negative orientation of 4-simplex $s^4_{\cW}$.
An element of the compact group $\Spin(4)$ and an element of the Clifford algebra
\begin{gather}
\Omega_{{\cV}_1{\cV}_2}=\Omega^{-1}_{{\cV}_2{\cV}_1}=\exp\left(
\omega_{{\cV}_1{\cV}_2}\right)=
\exp\left(\frac{1}{2}\sigma^{ab}
\omega^{ab}_{{\cV}_1{\cV}_2}\right)\in\Spin(4), 
\nonumber \\
\hat{e}_{{\cV}_1{\cV}_2}\equiv e^a_{{\cV}_1{\cV}_2}\gamma^a\equiv
-\Omega_{{\cV}_1{\cV}_2}\hat{e}_{{\cV}_2{\cV}_1}\Omega_{{\cV}_1{\cV}_2}^{-1},
\nonumber \\
|e_{{\cV}_1{\cV}_2}|<1, \quad |e_{{\cV}_1{\cV}_2}|\equiv\sqrt{\sum_a(e^a_{{\cV}_1{\cV}_2})^2}
\label{Variables_Grav}
\end{gather}
are assigned for each oriented 1-simplex $a_{{\cV}_1}a_{{\cV}_2}$.
The boundedness of the tetrad according to (\ref{Variables_Grav}) is necessary for the convergence of the functional integral of the partition function. By this limitation imposed on the tetrad, the present work differs from the author’s previous works on the theory of diskette gravity.
The conjecture is that the set of variables $\{\Omega,\,\hat{e}\}$  is an independent set of dynamic variables.
Fermionic degrees of freedom (Dirac spinors) are assigned to each vertex of the complex:
\begin{gather}
\Psi^{\dag}_{\cV}, \quad \Psi_{\cV}.
\label{Variables_Ferm}
\end{gather}
The set of variables $\{\Psi^{\dag},\,\Psi\}$ is a set of mutually independent variables, and the spinors $\Psi^{\dag}_{\cV}$ and $\Psi_{\cV}$ are in mutual involution (or anti-involution) relative to Hermitian conjugation operation.

Consider a model with an action
\begin{gather}
\mA=\mA_g+\mA_{\Psi}+\mA_{\Lambda_0}.
\label{Action_4D}
\end{gather}
Here $\mA_g$ and $\mA_{\Psi}$ are the actions of pure gravity and Dirac field, correspondingly:
\begin{gather}
\mA_g=-\frac{1}{5!\cdot2\cdot l_P'^2}\sum_{\cW}\sum_{\sigma}
\varepsilon_{\sigma({\cV}_{({\cW})1})\sigma({\cV}_{({\cW})2})
\sigma({\cV}_{({\cW})3})\sigma({\cV}_{({\cW})4})\sigma({\cV}_{({\cW})5})}
\nonumber \\
\times\tr\gamma^5\bigg\{
\Omega_{\sigma({\cV}_{({\cW})5})\sigma({\cV}_{({\cW})1})}
\Omega_{\sigma({\cV}_{({\cW})1})\sigma({\cV}_{({\cW})2})}\Omega_{\sigma({\cV}_{({\cW})2})\sigma({\cV}_{({\cW})5})}
\hat{e}_{\sigma({\cV}_{({\cW})5})\sigma({\cV}_{({\cW})3})}
\hat{e}_{\sigma({\cV}_{({\cW})5})\sigma({\cV}_{({\cW})4})}\bigg\}.
\label{Latt_Action_Grav}
\end{gather}
Each $\sigma$ is one of 5! vertex permutations ${\cV}_{({\cW})i}\longrightarrow\sigma(
{\cV}_{({\cW})i})$.
\begin{gather}
\mA_{\Psi}=\frac{1}{5\cdot24^2}\sum_{\cW}\sum_{\sigma}
\varepsilon_{\sigma({\cV}_{({\cW})1})\sigma({\cV}_{({\cW})2})
\sigma({\cV}_{({\cW})3})\sigma({\cV}_{({\cW})4})\sigma({\cV}_{({\cW})5})}
\nonumber \\
\times\tr\gamma^5\bigg\{ \hat{\Theta}_{\sigma({\cV}_{({\cW})5})\sigma({\cV}_{({\cW})1})}
\hat{e}_{\sigma({\cV}_{({\cW})5})\sigma({\cV}_{({\cW})2})}
\hat{e}_{\sigma({\cV}_{({\cW})5})\sigma({\cV}_{({\cW})3})}
\hat{e}_{\sigma({\cV}_{({\cW})5})\sigma({\cV}_{({\cW})4})}\bigg\},
\label{Latt_Action_Ferm}
\end{gather}
\begin{gather}
\hat{\Theta}_{{\cV}_1{\cV}_2}
\equiv\Theta^a_{{\cV}_1{\cV}_2}\gamma^a=\hat{\Theta}_{{\cV}_1{\cV}_2}^{\dag},  \quad
\Theta^a_{\cV_1\cV_2}=\frac{i}{2}\left(\Psi^{\dag}_{\cV_1}\gamma^a\Omega_{{\cV}_1{\cV}_2}\Psi_{\cV_2}-
\Psi^{\dag}_{\cV_2}\Omega_{{\cV}_2{\cV}_1}\gamma^a\Psi_{\cV_1}\right).
\label{Dirac_Form}
\end{gather}
 It is easy to check that (compare with (\ref{Variables_Grav}))
\begin{gather}
\hat{\Theta}_{{\cV}_1{\cV}_2}
\equiv-\Omega_{{\cV}_1{\cV}_2}\hat{\Theta}_{{\cV}_2{\cV}_1}
\Omega_{{\cV}_1{\cV}_2}^{-1}.
\label{Dir_Bil_Form_Trans}
\end{gather}
The contribution to the lattice action from the cosmological constant has the form
\begin{gather}
\mA_{\Lambda_0}=-\frac{1}{5!\cdot12}\cdot\frac{\Lambda_0}{l_P'^2}\varepsilon_{abcd}\sum_{\cW}\sum_{\sigma}
\varepsilon_{\sigma({\cV}_{({\cW})1})\sigma({\cV}_{({\cW})2})
\sigma({\cV}_{({\cW})3})\sigma({\cV}_{({\cW})4})\sigma({\cV}_{({\cW})5})}
\nonumber \\
\times e^a_{\sigma({\cV}_{({\cW})5})\sigma({\cV}_{({\cW})1})}
e^b_{\sigma({\cV}_{({\cW})5})\sigma({\cV}_{({\cW})2})}
e^c_{\sigma({\cV}_{({\cW})5})\sigma({\cV}_{({\cW})3})}
e^d_{\sigma({\cV}_{({\cW})5})\sigma({\cV}_{({\cW})4})}.
\label{Latt_Action_Lambda}
\end{gather}
The partition function is represented by the integral
\begin{gather}
Z=\prod_{1-\mbox{simplices}}\int_{|e_{{\cV}_1{\cV}_2}|<1}\prod_a\d e^a_{\cV_1\cV_2}
\int\d\mu\{\Omega_{\cV_1\cV_2}\}\prod_{\cV}\int\d\Psi^{\dag}_{\cV}\d\Psi_{\cV}\exp(\mA).
\label{Partition_function}
\end{gather}
Everywhere $\d\mu\{\Omega_{{\cV}_1{\cV}_2}\}$ is an invariant measure on the group $\Spin(4)$.

The action (\ref{Action_4D}) as well as integral (\ref{Partition_function}) are invariant relative to the gauge transformations
\begin{gather}
\tilde{\Omega}_{{\cV}_1{\cV}_2}
=S_{{\cV}_1}\Omega_{{\cV}_1{\cV}_2}S^{-1}_{{\cV}_2}, \quad
\tilde{\hat{e}}_{{\cV}_1{\cV}_2}=S_{{\cV}_1}\,\hat{e}_{{\cV}_1{\cV}_2}\,S^{-1}_{{\cV}_1}, \quad
\tilde{\Psi}_{\cV}=S_{\cV}\Psi_{\cV}, \quad \tilde{\Psi^{\dag}}_{\cV}=\Psi_{\cV}^{\dag}S_{\cV}^{-1}, \quad  S_{\cV}\in\Spin(4).
\label{Gauge_Trans}
\end{gather}
Verification of this fact is facilitated by using the relation (compare with the relation for
$\hat{e}_{{\cV}_1{\cV}_2}$ in (\ref{Gauge_Trans}))
\begin{gather}
\tilde{\hat{\Theta}}_{{\cV}_1{\cV}_2}=S_{{\cV}_1}\hat{\Theta}_{{\cV}_1{\cV}_2}S^{-1}_{{\cV}_1},
\label{Teta_Gauge_Trans}
\end{gather}
which follows directly from (\ref{Gauge_Trans}).

The considered lattice model is invariant with respect to the global discrete $\Z_2$ symmetry, which is an analog of the combined PT-symmetry. Let $\hat{\cal U}_{PT}$ denote the operator of this transformation. Then the transformed dynamic variables are expressed in terms of the original variables as follows:
\begin{gather}
\hat{\cal U}_{PT}^{-1}\Psi_{\cV}\hat{\cal U}_{PT}=U_{PT}\left(\Psi^{\dag}_{\cV}\right)^t, \quad
\hat{\cal U}_{PT}^{-1}\Psi^{{\dag}}_{\cV}\hat{\cal U}_{PT}=-\left(\Psi_{\cV}\right)^tU^{-1}_{PT}, \quad U_{PT}=i\gamma^1\gamma^3
\nonumber \\
\hat{\cal U}_{PT}^{-1}e^a_{{\cV}_1{\cV}_2}\hat{\cal U}_{PT}=-e^{a}_{{\cV}_1{\cV}_2}, \quad
\hat{\cal U}_{PT}^{-1}\omega^{ab}_{{\cV}_1{\cV}_2}\hat{\cal U}_{PT}=\omega^{ab}_{{\cV}_1{\cV}_2}.
\label{PT_transform}
\end{gather}
Here the superscript $"t"$ denotes the matrix transposition of the Dirac matrices and spinors.
We have:
\begin{gather}
U^{-1}_{PT}\gamma^aU_{PT}=(\gamma^a)^t, \quad
U^{-1}_{PT}\sigma^{ab}U_{PT}=-(\sigma^{ab})^t.
\label{PT_trans_Dir_Alg}
\end{gather}
It follows from (\ref{PT_transform}) and (\ref{PT_trans_Dir_Alg}) that
\begin{gather}
 U^{-1}_{PT}\Omega_{{\cV}_1{\cV}_2}U_{PT}=\left(\Omega_{{\cV}_2{\cV}_1}\right)^t,
\label{PT_trans_Conn}
\end{gather}
as well as
\begin{gather}
\left(\Theta^a_{\cV_1\cV_2}\right)^{PT}=-\Theta^a_{\cV_1\cV_2}.
\label{PT_trans_Ferm_Bil_Fjrm}
\end{gather}

\subsection{The long-wavelength limit}

Let's pass on to the long-wave limit, that is, to the limit of fields that slowly change when moving along the lattice. In this limit, the action (\ref{Action_4D}) transforms into the well-known continuum Palatini action plus the contribution of the cosmological constant, and the Dirac field turns out to be minimally coupled to gravity. This transformation have meaning
together with the transition to Minkowski signature. As a result the compact gauge group $\Spin(4)$ transforms into the non-compact group $\Spin(3,1)$.
In this section, all lattice quantities for the Euclidean signature are primed. For the quantities related to the Minkowski signature, the previous designations are retained.

For the indicated transformation of the action, the following deformations of the integration contours in the integral (\ref{Partition_function}) are necessary:
\begin{gather}
{\omega'}_{{\cV}_1{\cV}_2}^{4\alpha}= i\omega^{0\alpha}_{{\cV}_1{\cV}_2}, \quad
{\omega'}_{{\cV}_1{\cV}_2}^{\alpha\beta}=-\omega_{{\cV}_1{\cV}_2}^{\alpha\beta},
\nonumber \\
{e'}^4_{{\cV}_1{\cV}_2}= e^0_{{\cV}_1{\cV}_2}, \quad {e'}^{\alpha}_{{\cV}_1{\cV}_2}= ie^{\alpha}_{{\cV}_1{\cV}_2}.
\label{Variables_Trans_Mink}
\end{gather}
The variables $\omega^{ab}_{{\cW}ij}$, $e^a_{{\cW}ij}$ are real quantities for Minkowski signature,
and the indices take on the values
\begin{gather}
a,\,b\ldots=0,1,2,3, \quad \alpha,\,\beta,\ldots=1,2,3.
\label{Mink_Ind}
\end{gather}
The Dirac matrices are transformed as follows:
\begin{gather}
{\gamma'}^4=\gamma^0, \quad {\gamma'}^{\alpha}= i\gamma^{\alpha}, \quad
{\gamma'}^5=\gamma^5=i\gamma^0\gamma^1\gamma^2\gamma^3,
\nonumber \\
\frac12(\gamma^a\gamma^b+\gamma^b\gamma^a)=\eta^{ab}=\diag(1,\,-1,\,-1,\,-1), \quad
\tr\gamma^5\gamma_a\gamma_b\gamma_c\gamma_d=4i\varepsilon_{abcd}, \quad \varepsilon_{0123}=1.
\label{Mink_Dirac_matr}
\end{gather}
Thus, for $\sigma^{ab}=(1/4)[\gamma^a,\,\gamma^b]$ we get
\begin{gather}
{\sigma'}^{4\alpha}= i\sigma^{0\alpha}, \quad {\sigma'}^{\alpha\beta}=-\sigma^{\alpha\beta}.
\label{Mink_Spin_Matr}
\end{gather}
Raising and lowering indices $a,b,\ldots$ is done using tensors $\eta^{ab}$ and $\eta_{ab}$, respectively.
 As a result of (\ref{Variables_Trans_Mink})-(\ref{Mink_Spin_Matr}) we have
\begin{gather}
{\omega'}_{{\cV}_1{\cV}_2}=\frac12\omega^{ab}_{{\cV}_1{\cV}_2}\,\sigma_{ab}
\equiv\omega_{{\cV}_1{\cV}_2},
\quad
{\hat{e}'}_{{\cV}_1{\cV}_2}=\gamma_ae^a_{{\cV}_1{\cV}_2}\equiv\hat{e}_{{\cV}_1{\cV}_2},
\label{Mink_5}
\end{gather}
and also
\begin{gather}
{\Omega'}_{{\cV}_1{\cV}_2}=\exp\left(\frac12{\omega'}_{{\cV}_1{\cV}_2}^{ab}{\sigma'}^{ab}\right)=
\exp\left(\frac12\omega_{{\cV}_1{\cV}_2}^{ab}
\sigma_{ab}\right)\equiv\Omega_{{\cV}_1{\cV}_2}\in\Spin(3,1).
\label{Mink_6}
\end{gather}
We see that the holonomy elements $\Omega_{{\cV}_1{\cV}_2}$ become the elements of the non-compact group $\Spin(3,1)$.

Dirac variables are transformed according to
\begin{gather}
\Psi'_{\cV}=\Psi_{\cV}, \quad  {\Psi'}^{\dag}_{\cV}=\Psi_{\cV}^{\dag}\gamma^0=\overline{\Psi}_{\cV}.
\label{Mink_Dir}
\end{gather}
in passing to the Minkowski signature.

The transition to the long-wave limit is possible for such field configurations that change quite slowly during transitions from simplex to simplex, that is, with small or significant movements along the lattice. This rule applies to any grating. In our theory, it is precisely at the stage of transition to the long-wave limit that the need to introduce local coordinates arises. Local coordinates are the markers of the lattice vertices. Consider some 4D subcomplex ${\mK}'\in\mK$ with the trivial topology of a four-dimensional disk and geometric realization in $\R^4$. Thus, each vertex of the subcomplex acquires coordinates $x^{\mu}$, which are the coordinates of the image of the vertex in $\R^4$:
\begin{gather}
x^{\mu}_{\cV}\equiv x^{\mu}(a_{\cV}),
 \qquad \ \mu=1,\,2,\,3,\,4.
\label{intr110}
\end{gather}
At this stage, the coordinates are dimensionless, that is, real numbers. Let's consider some simplex $s^4_{\cW}\in{\mK}'$. Let us denote all five vertices of this 4-simplex as ${\cV}_i$, $i=1,2,3,4$ and ${\cV}_m\neq {\cV}_i$. The properties of the geometric realization are such that four infinitesimal vectors
\begin{gather}
\d x^{\mu}_{{\cV}_m{\cV}_i}\equiv x^{\mu}_{{\cV}_i}-x^{\mu}_{{\cV}_m}=
-\d x^{\mu}_{{\cV}_i{\cV}_m}\in\R^4 ,  \quad
i=1,\,2,\,3,\,4
\label{intr120}
\end{gather}
are linearly independent. 

In the works
\cite{vergeles2006one,vergeles2015wilson,vergeles2017fermion,vergeles2017note,vergeles2021note} it is proven that there are 1-forms in $\R^4$
$\omega_{\mu}(x)$ and $\hat{e}_{\mu}(x)$ such that the equalities 
\begin{gather}
\omega_{\mu}\left(\frac{1}{2}\,(x_{{\cV}_m}+
x_{{\cV}_i})\,\right)\d
x^{\mu}_{{\cV}_m{\cV}_i}=\omega_{{\cV}_m{\cV}_i},
\label{intr160}
\end{gather}
\begin{gather}
\hat{e}_{\mu}\left(\frac{1}{2}\,(x_{{\cV}_m}+
x_{{\cV}_i})\,\right)\d x^{\mu}_{{\cV}_m{\cV}_i}=
\hat{e}_{{\cV}_m{\cV}_i}.
\label{intr190}
\end{gather}
are true. It is further assumed that the 1-forms smoothly depend on points in $\R^4$ that are images of the vertices of the complex. In the long-wave limit, the elements $\Omega_{{\cV}_m{\cV}_i}$ are close to unity and, up to $O\big((\d x)^2\big)$ inclusive, the formula 
\begin{gather}
\Omega_{{\cV}_m{\cV}_i}\Omega_{{\cV}_i{\cV}_j}\Omega_{{\cV}_j{\cV}_m}=
\exp\left[\frac{1}{2}\,\mR_{\mu\nu}(x_{{\cV}_m})\d x^{\mu}_{\left({\cV}_m{\cV}_i\right)} \d
x^{\nu}_{\left({\cV}_m{\cV}_j\right)}\right]
\label{Lat_curv}
\end{gather}
holds, where
\begin{gather}
\mR_{\mu\nu}=\partial_{\mu}\omega_{\nu}-\partial_{\nu}\omega_{\mu}+
[\omega_{\mu},\,\omega_{\nu}\,]\,.
\label{Long_W_Curv}
\end{gather}

Let's write down the long-wave  limit of the action (\ref{Action_4D}):
\begin{gather}
{\mA'}_{g\,Lat}\longrightarrow i\mA_g,  \quad \mA_g=-\frac{1}{4\,l^2_P}\varepsilon_{abcd}\int\mR^{ab}\wedge e^c\wedge e^d,  \quad
\mR^{ab}=\mR^{ab}_{\mu\nu}\d x^{\mu}\wedge\d x^{\nu},
\label{Long_Wav_Grav_Act}
\end{gather}
\begin{gather}
{\mA'}_{\Psi\,Lat}\longrightarrow i\mA_{\Psi}, \quad \mA_{\Psi}=\frac16\varepsilon_{abcd}\int\Theta^a\wedge e^b\wedge e^c\wedge e^d,
\nonumber \\
\Theta^a=\frac{i}{2}\left[\overline{\Psi}\gamma^a{\cal D}_{\mu}\,\Psi-
\left(\overline{{\cal D}_{\mu}\,\Psi}\right)\gamma^a\Psi\right]\d x^{\mu}, \quad
{\cal D}_{\mu}=\left(\partial_{\mu}+\omega_{\mu}\right),
\label{Long_Wav_Dir_Act}
\end{gather}
\begin{gather}
\mA'_{\Lambda_0\,Lat}\longrightarrow i\mA_{\Lambda_0}, \quad
\mA_{\Lambda_0}=-\frac{2\Lambda_0}{l_P^2}\int e^0\wedge e^1\wedge e^2\wedge e^3.
\label{Long_Wav_Lambda_Act}
\end{gather}
All other terms in such a transition will contain additional factors in the positive power $(l_P/\lambda)\longrightarrow0$, and therefore they are omitted. Here $\lambda$ is the characteristic wavelength of the physical subsystem. This situation is typical when going to the long-wave limit in any lattice theory.

For clarity, let us point out the fact that on the lattice all variables and constants are dimensionless and of order one. In particular, the constant $l_P'\sim1$ in (\ref{Latt_Action_Grav}) and (\ref{Latt_Action_Lambda}) is dimensionless, as are the differentials $\d x^{\mu}_{{\cV}_m{\cV} _i}$ in (\ref{intr120}). When moving to dimensional quantities, we assume
\begin{gather}
\d x^{\mu}_{{\cV}_m{\cV}_i}=\d x^{\mu}/l_P\sim1,
 \label{Long_Wav_Coord}
\end{gather}
where the differential $\d x^{\mu}$ is measured in centimeters and $l_P\sim10^{-32}\mbox{cm}$ (see (\ref{Estimation_1})). From (\ref{Long_Wav_Coord}) it is clear that the step of the irregular lattice has a size of the order of $l_P$. And at the same time, all components of the action are dimensionless, but variables and constants acquire dimension. For example, in the term (\ref{Long_Wav_Lambda_Act}) the cosmological constant is $\Lambda_0\sim l_P^{-2}$.

The action (\ref{Long_Wav_Grav_Act})-(\ref{Long_Wav_Lambda_Act}) is a Hilbert-Einstein action, minimally related to the Dirac field and written in Palatini form. It is invariant under diffeomorphisms. This fact is not accidental, since in (\ref{intr110}) the very method of introducing coordinates is such that already at this stage the independence of the action from the arbitrariness of introduced coordinates is visible. 
It is important that small terms in the long-wave limit, proportional to positive powers of $(l_P/\lambda)$, are also invariant under diffeomorphisms. Indeed, when expanding the action (\ref{Action_4D}) in terms of the parameter $\l_P$, which does not depend on the coordinates, all the coefficients at the powers of this parameter will also be independent on the coordinates.

In the Minkowski signature, the PT-symmetry of the action is determined by the formulas (\ref{PT_transform})-(\ref{PT_trans_Dir_Alg}), (\ref{PT_trans_Ferm_Bil_Fjrm}) with the only difference that in these formulas the replacement should be made
$\Psi^{\dag}\longrightarrow\overline{\Psi}$.

\section{Superhigh temperatures in the lattice model of gravity}

Let us consider the state of lattice theory at ultrahigh temperatures.
As is known, in quantum theory, the partition function $\tr\exp(-\beta\cH)$ differs from the transition amplitude over a finite time $\Delta t$, represented as a functional integral, by the transition to imaginary time
$\Delta t=-i\beta$, $\beta=1/T$,  and taking the trace. We must study some properties of the partition function (\ref{Partition_function}) in the case of ultra-high temperatures, that is, $\beta\longrightarrow0$.

When applied to the lattice theory studied here, this means the following.

Suppose that a 4D lattice has two 3D sublattices $\Sigma_1$ and $\Sigma_2$ that form its boundary. For simplicity, we assume that between $\Sigma_1$ and $\Sigma_2$ there are $N<\infty$ 4D lattice layers, each one edge thick. The sublattices $\Sigma_1$ and $\Sigma_2$ are considered identical, that is, there is a one-to-one correspondence between all elements of these sublattices. The last property makes it possible to calculate the partition function.

Let $\Phi_{1\xi}$ and $\Phi_{2\xi}$ denote the holomorphic functions of fermion variables with real coefficients defined on $\Sigma_1$ and $\Sigma_2$, respectively.
To calculate the trace in fermionic variables, it is necessary to use the full set of holomorphic wave functions $\Phi_{1\xi}\equiv\Phi_{1\xi}\{\Psi^{\dag}_{{\cV}}\}$, and their Hermitian conjugate functions $\Phi_{2\xi}^{\dag}\equiv\left(\Phi_{2\xi}\{\Psi^{\dag}_{{\cV}}\}
\right)^{\dag}=\Phi_{2\xi}\{\Psi_{{\cV}}\}$.
Here the index $\xi$ lists the independent wave functions from their complete set.  The functional
\begin{gather}
\Phi_{2\xi}^{\dag}\exp(\beta\mA)\Phi_{1\xi}
\label{product}
\end{gather}
should be placed under the integral (\ref{Partition_function}) and the sum over $\xi$ taken.
In the integral (\ref{Partition_function}) on the identified 1-simplices belonging to $\Sigma_1$ and $\Sigma_2$, the variables $\{\Omega\}$ and $\{e^a\}$ must be identified .

We are interested in the case $\beta\ll1$.
We intend to prove that discrete PT symmetry (\ref{PT_transform})-(\ref{PT_trans_Ferm_Bil_Fjrm}) is not broken at ultrahigh temperatures.

The following Statement holds: \\

{\it Statement 1}. In a certain finite neighborhood of the point $\beta=0$, the free energy of the partition function (\ref{Partition_function}), with the exception of the term of the form $(-(C_1\cdot\mM+4\mN\nu_{\Psi})\cdot\ln\beta)$ (see (\ref{Part_Func_Hol})), is a holomorphic function of the variable $\beta$. All  symmetries of the action (\ref{Action_4D}), including discrete PT symmetries, are preserved. $\square$ \\

Let us give some arguments in favor of the Statement 1. Let us consider the high-temperature expansion of the partition function in the 2D Ising model, which is the sum over all closed paths with self-intersections on the lattice. Let the lattice contain $\mL\longrightarrow\infty$ edges and let a path (possibly several non-intersecting paths) with a total of $\ml$ edges be fixed on it. Note that $\ml$ is an even number and $\ml\geq4$. When calculating the partition function, it is convenient to isolate the factor $(\cosh\beta)^{\mL}$, which does not give a singularity in the free energy. Then each edge of the path should be assigned a weight $(\tanh\beta)$. Let $g_{\ml}$ be the number of such closed paths. We have a strict inequality
\begin{gather}
g_{\ml}<\frac{\mL!}{\ml!(\mL-\ml)!},
\label{Number_of_Paths}
\end{gather}
since the number on the right side of the inequality (\ref{Number_of_Paths}) also includes the number of all non-closed paths of weight $(\tanh\beta)^{\ml}$. Therefore, the Ising partition function $Z$
\begin{gather}
Z/(\cosh\beta)^{\mL}=1+\sum_{\ml}g_{\ml}(\tanh\beta)^{\ml}<  \sum_{\ml=0}^{\mL}\frac{\mL!}{\ml!(\mL-\ml)!}(\tanh\beta)^{\ml}
\nonumber \\
=(1+\tanh\beta)^{\mL}\equiv Z_M.
\label{P_F_MAJOR}
\end{gather}
This shows that the specific free energy per edge $f_M=-\ln Z_M/\mL=-\ln(1+\tanh\beta)$ is a holomorphic function in a finite neighborhood of the point $\beta=0$ and in the case of $\mL\longrightarrow\infty$. This observation is a consequence of the fact that the quantity $Z_M$ is essentially local for small $\beta$, that is, effectively $Z_M$ is the product of local holomorphic functions. In other words, there is no long-range order at small $\beta$. But if open contours are eliminated from the sum $Z_M$, then long-range order will not arise.
However, the size of the holomorphy neighborhood may change. This means that the limit $f=-\ln Z/\mL$ at $\mL\longrightarrow\infty$ exists. Indeed, the value on the left side (\ref{P_F_MAJOR}) can be represented as
\begin{gather}
Z/(\cosh\beta)^{\mL}=[1+\rho(\tanh\beta)]^{\mL},
\label{Holomorphic_Z}
\end{gather}
and the holomorphic function $\rho(\tanh\beta)\rightarrow0$ for $\beta\rightarrow0$. That's why
\begin{gather}
f=-\ln\cosh\beta-\ln[1+\rho(\tanh\beta)]
\label{Holomorphic_F}
\end{gather}
is a holomorphic function in some finite neighborhood of the point $\beta=0$.

Similar reasoning can be carried out in the case of a partition function (\ref{Partition_function}), although this case is qualitatively more complicated.
Both the fermion and bosonic integrals in (\ref{Partition_function}) are well defined. Let us consider qualitatively the results of these integrations.

Let ${\mN}\longrightarrow\infty$ and $\nu_{\Psi}$ denote the numbers of vertices of the complex and Dirac fields, respectively. Integration into (\ref{Partition_function}) over the Dirac field gives the factor under the remaining integrals of the form
\begin{gather}
\beta^{4{\mN}\nu_{\Psi}}{\mP}\{\Omega,e^a\},
\label{Dirac_Polynom}
\end{gather}
where ${\mP}\{\Omega,e^a\}$ is a gauge-invariant homogeneous polynomial of the variables $\Omega$ and $e^a$ of degrees ${4{\mN}\nu_{\Psi}}$ and ${12{\mN}\nu_{\Psi}}$, respectively.

Next, we consider integrals over bosonic variables. First we consider integration over the connection variables $\Omega$.

Let the polynomial ${\mP}\{\Omega,e^a\}$ contain a variable $\Omega_{{\cV}_1{\cV}_2}$  to an odd power. Integral $(\int\d\mu\{\Omega_{{\cV}_1{\cV}_2}\}\ldots)$ of the tensor product of an odd number of elements  $\Omega_{{\cV}_1{\cV }_2}$ is equal to zero. Let us expand the exponential into terms containing the element $\Omega_{{\cV}_1{\cV}_2}$. Let us assume that as a result of a finite number of such steps (for a finite subcomplex) we will have a tensor product of an even number of elements $\Omega_{{\cV}_1{\cV}_2}$ on each 1-simplex $a_{{\cV}_1}a_{{\cV}_2}$.
Therefore, as a result of integration over the gauge group, the first nonzero term will receive an additional factor $\beta^{C\cdot\mM}$, where $\mM\rightarrow\infty$ is the number of 1-simplices of the complex and $0\leq C\sim1$. All other terms of this expansion, as a result of integration over the gauge group, give an additional gauge invariant term $\mF\{e^a;\beta\}$ in the partition function. This term is a functional of the variables $\{e^a\}$ and a well-converging holomorphic function of the parameter $\beta$ in the neighborhood of zero, and
\begin{gather}
\frac{\mF\{e^a;\beta\}}{\beta^{C\cdot\mM+4\mN\nu_{\Psi}}}\rightarrow0 \quad \mbox{as} \quad
\beta\rightarrow0.
\nonumber
\end{gather}
There are only two types of invariants relative to gauge transformations (\ref{Gauge_Trans}) associated with a vertex $a_{\cV}$:
\begin{gather}
e^a_{{\cV}{\cV_1}}e^a_{{\cV}{\cV_2}},
\label{O(4)Invariants_I}
\end{gather}
\begin{gather}
\varepsilon_{abcd}e^a_{{\cV}{\cV_1}}e^b_{{\cV}{\cV_2}}e^c_{{\cV}{\cV_3}}e^d_{{\cV}{\cV_4}}.
\label{O(4)Invariants_II}
\end{gather}
In (\ref{O(4)Invariants_I}) the vertices $a_{{\cV}_1}$ and $a_{{\cV}_2}$ are not necessarily different, and in (\ref{O(4)Invariants_II}) the 1-simplices $(a_{\cV}a_{{\cV}_1})$,
$(a_{\cV}a_{{\cV}_2})$, $(a_{\cV}a_{{\cV}_3})$, $(a_{\cV}a_{{\cV}_4})$ do not necessarily belong to the same 4-simplex.

Let us briefly consider the integral over variables $\{e^a\}$. Note that if under the integral there is an expression (\ref{O(4)Invariants_I}) to the first degree and it does not intersect with other similar expressions, then the integral is identically equal to zero. A similar statement holds for the expression (\ref{O(4)Invariants_II}). Therefore, under the integral the expressions (\ref{O(4)Invariants_I}) and (\ref{O(4)Invariants_II})  must be in such powers and combinations that the variables $e^a_{{\cV}_1{\cV}_2}$ on each 1-simplex  are in an even power (including zero). For this purpose, expansion of the exponential in (\ref{product}) in the contribution to the action $\beta\mA_{\Lambda_0}$ (see (\ref{Latt_Action_Lambda})) may be necessary.

The action (\ref{Action_4D}) is local, that is, it consists of terms (local operators), each of which is defined on the nearest lattice elements. Let's call one of these operators $A$, and the other $B$, and we will assume that the operators $A$ and $B$ are separated by a considerable distance along the lattice. Consider the correlator $\langle AB\rangle$.
For this correlator to be non-zero, it is necessary to expand the exponential in these operators so that between the operators $A$ and $B$ there is a sufficient number of adjacent local operators. But such an expansion will give the parameter $\beta\ll1$ to a fairly high degree, at least proportional to the number of 1-simplices between $A$ and $B$, and with a coefficient of the order of unity. Therefore, the partition function in the high-temperature phase is local in nature.

Let us draw a conclusion from the above consideration. At ultrahigh temperature $(\beta=1/T\longrightarrow0)$ the integral for the partition function (\ref{Partition_function})  is a holomorphic function in a neighborhood of the point $\beta=0$ and has the form (compare with (\ref{Holomorphic_Z}))
\begin{gather}
Z=\Const\cdot\beta^{C_1\cdot\mM+4\mN\nu_{\Psi}}(1+\mf(\beta))^{C_2\cdot\mM},
\nonumber \\
\mf(\beta)\rightarrow0 \quad \mbox{as} \quad \beta\rightarrow0, \quad C_1\sim C_2\sim1.
\label{Part_Func_Hol}
\end{gather}
This implies the validity of the Statement 1. In turn, it follows from the Statement1 that in the high-temperature phase no symmetries are broken, including discrete PT symmetry.

Based on the validity of the Statement 1, we will begin to calculate the averages of some operators.

The discrete PT transformation (\ref{PT_transform})-(\ref{PT_trans_Ferm_Bil_Fjrm}) does not change the measure and integrand (except, perhaps, the averaged value), but swaps the initial and final states:
\begin{gather}
\left(\Phi_{2\xi}^{\dag}\right)'\equiv\Phi_{2\xi}^{\dag}\hat{\cal U}_{PT}^{\dag}=\Phi_{2\xi}\{U_{PT}(\Psi_{\cV}^{\dag})^t\},
\nonumber \\
\Phi_{1\xi}'\equiv\hat{\cal U}_{PT}\Phi_{1\xi}=\Phi_{1\xi}
\{-\Psi^t_{\cV}U^{-1}_{PT}\},
\nonumber \\
\left(\Phi_{2\xi}^{\dag}\right)'\cdot\Phi_{1\xi}'=\Phi_{1\xi}
\{\Psi^t_{\cV}U^{-1}_{PT}\}\cdot\Phi_{2\xi}\{U_{PT}(\Psi_{\cV}^{\dag})^t\}.
\label{PT_Trans_Fermi_St}
\end{gather}
Here, a prime above the wave function symbol means that it is PT-transformed. From the last equality it is clear that the initial and final wave functions are swapped as a result of the PT transformation, and their scalar product is preserved. This means that the operator $\hat{\cal U}_{PT}$ is anti-unitary.

According to the general rule, the integral does not change when the integration variables are changed. In our case, the integral over the fermionic and tetrad variables and the sum over the fermionic states $\xi$ are of interest. The integral over the connection variables $\{\Omega\}$ should be excluded. Otherwise, further reasoning will be meaningless. Let's consider two integrals.

First integral:
\begin{gather}
Z\{\Omega\}=\sum_{\xi}\int_e\int_{\Psi^{\dag}\Psi}
\Phi_{2\xi}^{\dag}\exp(\beta\mA)\Phi_{1\xi}\stackrel{PT}{\longrightarrow}Z\{\Omega\}.
\nonumber
\end{gather}

Second integral:
\begin{gather}
\langle{\Theta}^a_{{\cV}_1{\cV}_2}\rangle_{e,\Psi}=N\sum_{\xi}\int_e\int_{\Psi^{\dag}\Psi}
\Phi_{2\xi}^{\dag}{\Theta}^a_{{\cV}_1{\cV}_2}\exp(\beta\mA)\Phi_{1\xi} \quad
\stackrel{PT}{\longrightarrow}\quad-\langle{\Theta}^a_{{\cV}_1{\cV}_2}\rangle_{e,\Psi}.
\nonumber
\end{gather}

Let's consider the second integral in more detail. We have a chain of equalities:
\begin{gather}
\langle{\Theta}^a_{{\cV}_1{\cV}_2}\rangle_{e,\Psi}=N\sum_{\xi}\int_e\int_{\Psi^{\dag}\Psi}
\Phi_{2\xi}^{\dag}{\Theta}^a_{{\cV}_1{\cV}_2}\exp(\beta\mA)\Phi_{1\xi}
\nonumber \\
=N\sum_{\xi}\int_e\int_{\Psi^{\dag}\Psi}\left(\Phi_{2\xi}^{\dag}\right)'
{\Theta}^a_{{\cV}_1{\cV}_2}\exp(\beta\mA)\Phi_{1\xi}'
\nonumber \\
=N\sum_{\xi}\int_e\int_{\Psi^{\dag}\Psi}
\Phi_{1\xi}^{\dag}\left\{\hat{\cal U}_{PT}^{-1}{\Theta}^a_{{\cV}_1{\cV}_2}\exp(\beta\mA)
\hat{\cal U}_{PT}\right\}^{\dag}\Phi_{2\xi}
\nonumber \\
=-N\sum_{\xi}\int_e\int_{\Psi^{\dag}\Psi}
\Phi_{1\xi}^{\dag}{\Theta}^a_{{\cV}_1{\cV}_2}\exp(\beta\mA)\Phi_{2\xi}
=-\langle{\Theta}^a_{{\cV}_1{\cV}_2}\rangle_{e,\Psi}.
\label{Main_Result}
\end{gather}
Here $N\{\Omega\}$ is a normalizing constant, the integral is calculated only over the tetrad and Dirac field variables, but not over the connection variables. If the integration also included an integral over the variables $\{\Omega\}$, then any gauge-non-invariant quantity under the integral, like ${\Theta}^a_{{\cV}_1{\cV}_2}$, would turn into zero identically. But the integral (\ref{Main_Result}) is meaningful, since the gauge is fixed in it. In (\ref{Main_Result}), the first equality is the definition of the average value, the second equality is a consequence of the Statement 1, the third and fourth equalities are consequences of the equalities (\ref{PT_Trans_Fermi_St}) and (\ref{PT_trans_Ferm_Bil_Fjrm}), respectively.

The main conclusion of this work follows from the chain of equalities (\ref{Main_Result}):
\begin{gather}
\langle{\Theta}^a_{{\cV}_1{\cV}_2}\rangle_{e,\Psi}\equiv
\langle{\Theta}^a_{{\cV}_1{\cV}_2}\rangle_{\mbox{Gauge Fix}}=0.
\label{Mean_Zero}
\end{gather}

The vacuum mean of ${\Theta}^a_{{\cV}_1{\cV}_2}$ in the long-wave limit in the Minkowski signature at zero temperature was calculated in \cite{vergeles2021domain}:
\begin{gather}
\langle0|\Theta^a_{\mu}|0\rangle=\frac{2}{\pi^2l^4_P}e^a_{\mu}\neq0,
\nonumber \\
\Theta^a_{\mu}=\frac{i}{2}\left[\overline{\Psi}\gamma^a{\cal D}_{\mu}\,\Psi-
\left(\overline{{\cal D}_{\mu}\,\Psi}\right)\gamma^a\Psi\right], \quad
{\cal D}_{\mu}=\left(\partial_{\mu}+\omega_{\mu}\right).
\label{Theta_Mean_Minkov}
\end{gather}
A comparison of the equations (\ref{Mean_Zero}) and (\ref{Theta_Mean_Minkov}) shows that in the lattice theory of gravity coupled with the Dirac field, there is a phase transition in temperature.

The phase transition from a phase with a zero value of the average value $\Theta^a_{\mu}$ to a phase with a non-zero value of this value and the physical significance of this transition were considered in the work
\cite{volovik2022gravity}. The work of \cite{volovik1990superfluid} may also be interesting in this regard.

\section{Einstein equation and solution}

Here the standard spatially flat Robertson-Walker metric
\begin{gather}
\d s^2=\d t^2-a^2(t)(\d x^{\alpha})^2, \quad \alpha=1,2,3
\label{Metric}
\end{gather}
is used, $H={\dot a}/a$ is Hubble parameter and $a(t)$ is  the cosmic scale factor.
We assume $c=1$. Appropriate remarks are made in formulas with restored dimension.

First of all, we will show that at the end of the inflation process and further, vacuum quantum fluctuations are insignificant when describing the dynamics of macroscopic regions of space and the matter contained in them.

Our approach assumes that all physical quantities are determined by taking into account vacuum quantum zero point fluctuations. In particular, the energy density and pressure are mainly determined by quantum fluctuations.
It is well known  \cite{schwinger2018particles} that the following simultaneous commutation relations for the components of the energy-momentum tensor take place in the Minkowski space:
\begin{gather}
\left[T^{00}({\bf x}),T^{00}({\bf y})\right]=-i\hbar \left(T^{0k}({\bf x})+T^{0k}({\bf y})\right)
\partial_k\delta^{(3)}({\bf x}-{\bf y})+\mbox{Schwinger terms},
\label{En_mom_Tens_Cov}
\end{gather}
and so on. The Schwinger terms are the higher derivatives of the $\delta$-functions and are not of interest here. The same commutation relations take place in a curved space if they are written near the center of Riemann's normal coordinates.

Let us denote by $l_P$ the minimum possible size on the lattice, which we will call the Planck size. Then $\delta^{(3)}({\bf 0})\sim l_P^{-3}$. Thus we find, using (\ref{En_mom_Tens_Cov}) and the well-known rule, the order of quantum fluctuations of the energy density $\Delta\varepsilon$ at wavelengths $\lambda\gg l_P$:
\begin{gather}
\Delta\varepsilon\sim\sqrt{\frac{\hbar |\varepsilon|}{l_P^3\lambda}}, \quad
\frac{\Delta\varepsilon}{|\varepsilon|}\sim\sqrt{\frac{\hbar }{l_P^3\lambda|\varepsilon|}}.
\label{Fluc_En_Dens}
\end{gather}
In the modern era and also at the end of the inflation phase, the vacuum temperature $T_{\mbox{vac}}\sim\hbar H$ (see Section V) is very small compared to the maximum (modulo) particle energy in the Dirac sea $|\epsilon|\sim\hbar/l_P$. It follows that the vacuum energy density $|\varepsilon|\sim\hbar/l_P^4$, and we obtain the estimation
\begin{gather}
\Delta\varepsilon/|\varepsilon|\sim\sqrt{l_P/\lambda}\sim10^{-16} \quad \mbox{for}
\quad \lambda\sim1\,\mbox{cm}, \quad l_P\sim10^{-32}\,\mbox{cm}.
\label{Fluc_En_Dens_Now}
\end{gather}
Note that, provided that the long-wave limit is valid, it makes sense to assume $\lambda/a\sim\const_1$, $l_P/a\sim\const_2$, and therefore $l_P/\lambda\sim\const$. Therefore the assessment (\ref{Fluc_En_Dens_Now}) remains valid over a very wide range, including a significant part of the inflation phase.
This shows that the use of Einstein's classical equations to describe macroscopic dynamics in the modern era is justified.
However, when approaching the Big Bang point, the vacuum temperature becomes very high, as a result of which the energy density $|\varepsilon|$ quickly decreases in absolute value due to the population of positive energy levels by fermions. Estimates (\ref{Fluc_En_Dens}) and (\ref{Fluc_En_Dens_Now}) become unsatisfactory, and in the situation $\Delta\varepsilon/|\varepsilon|\sim1$ classical equations are unsuitable, the dynamics become purely quantum. A detailed study of the boundary of the transition from classical to quantum dynamics is not included in the scope of this work.

Let us demonstrate the tendency of the average energy of the system to zero with increasing temperature using a simple example. Consider one Fermi particle, which can be in only two states with energies $\pm\epsilon$. Suppose that the particle is placed in a thermostat, so its chemical potential is zero and the temperature $T$ tends to infinity ($\epsilon/T\longrightarrow0$). Then $\varepsilon=<\epsilon>\longrightarrow0$.

\subsection{Solution of Einstein's equations with bare cosmological constant}

We assume that the energy-momentum tensor of an ideal relativistic fluid is suitable for solving the problem posed:
\begin{gather}
T^a_b=(\varepsilon+p)U^aU_b-p\delta^a_b.
\label{Energy-momentum}
\end{gather}
We work in an orthonormal basis in which the metric tensor $\eta_{ab}=\diag(1,-1,-1,-1)$.
On the right side of (\ref{Energy-momentum}), the symbols $\varepsilon$ and $p$ denote the energy density and pressure, respectively, and these quantities also include vacuum energy and pressure. Since fermion fields, in contrast to bosonic ones, make a negative contribution to the vacuum energy, but there are significantly more fermionic degrees of freedom than bosonic ones, we have $\varepsilon<0$. Moreover, lattice regularization means that $|\varepsilon|,|p|<\infty$.
$U^a$ is the averaged 4-velocity of the macroscopic regions of the lattice. In our case $U^a=(1,0,0,0)$.
To compensate for the vacuum energy, a bare finite positive cosmological constant $\Lambda_0$ is introduced into the Einstein equation:
\begin{gather}
\mR^a_b-\frac12\delta^a_b\mR=8\pi G\,T^a_b+\Lambda_0\delta^a_b.
\label{Einstein_eq}
\end{gather}
In lattice theory, the cosmological constant is introduced in a natural way (see (\ref{Latt_Action_Lambda})).
We assume that the bare cosmological constant (with restored dimension)
\begin{gather}
\Lambda_0=\const\sim l_P^{-2}, \quad l_P\sim\sqrt{\frac{8\pi G\hbar}{c^3}}\sim10^{-32}\mbox{cm}.
\label{Estimation_1}
\end{gather}
Note that the bare cosmological constant on the lattice (\ref{Latt_Action_Lambda}) is dimensionless, like all other constants and variables, and it is of the order of unity.

For the metric, we use ansatz (\ref{Metric}).
In order not to clutter up the formulas,  we introduce the notation
\begin{gather}
8\pi G\,\varepsilon= \tilde{\varepsilon}, \quad
8\pi G\,p= \tilde{p}.
\label{Rescaling}
\end{gather}
All components of the Einstein equation are reduced to two equations:
\begin{gather}
3\frac{\dot{a}^2}{a^2}=\Lambda_0+\tilde{\varepsilon},  \quad
2\frac{\ddot{a}}{a}+\frac{\dot{a}^2}{a^2}=\Lambda_0-\tilde{p}.
\label{Einstein_eq_1}
\end{gather}
Another equation $\nabla_aT^a_b=0$ is a consequence of equations (\ref{Einstein_eq_1}), and therefore it does not need to be considered.  With the help of the Hubble parameter $H(t)$ Eqs. (\ref{Einstein_eq_1}) are rewritten as follows:
\begin{gather}
2\dot{H}+(\tilde{\varepsilon}+\tilde{p})=0,  \quad
3H^2-(\Lambda_0+\tilde{\varepsilon})=0.
\label{Einstein_eq_2}
\end{gather}
So we have 3 unknown functions $\{\tilde{\varepsilon}(t), \tilde{p}(t), H(t)\}$ and 2 equations (\ref{Einstein_eq_2}).
The missing equation is the equation of state relating energy density and pressure.
The following facts are known regarding the equation of state of relativistic matter: (i) in the case of real dust-like matter $p=0$; (ii) in the case of real ultrarelativistic matter $p=\varepsilon/3$; (iii) in the case of the energy density and vacuum pressure in de Sitter space $p=-\varepsilon$.
In all three cases, the energy density and pressure are linearly related. In addition, these quantities have the same dimensionality. Therefore, for the energy density and pressure, including vacuum values, we accept the following hypothesis of a linear relation:
\begin{gather}
\tilde{p}=\varkappa\Lambda_0+(\varkappa-1)\tilde{\varepsilon} \longleftrightarrow
\tilde{\varepsilon}+\tilde{p}=\varkappa(\tilde{\varepsilon}+\Lambda_0).
\label{State_equation}
\end{gather}
This equation is linear and inhomogeneous with an unknown dimensionless function $\varkappa(t)$, the asymptotics of which are further determined based on the known dynamics. The set of equations (\ref{Einstein_eq_2}) and (\ref{State_equation}) has a solution:
\begin{gather}
\dot{H}=-\frac32\varkappa H^2 \longrightarrow
\nonumber \\
H(t)=H_0\left(1+\frac32H_0\int_{t_0}^t\varkappa(t')\d t'\right)^{-1},
\label{Solution_H}
\end{gather}
\begin{gather}
\tilde{\varepsilon}(t)=-\Lambda_0+3H_0^2\left(1+\frac32H_0\int_{t_0}^t\varkappa(t')\d t'\right)^{-2},
\label{Solution_e}
\end{gather}
\begin{gather}
\tilde{p}(t)=\Lambda_0+3\big(\varkappa(t)-1\big)H_0^2\left(1+\frac32H_0\int_{t_0}^t
\varkappa(t')\d t'\right)^{-2}.
\label{Solution_p}
\end{gather}
Here $H_0$ is the integration constant which plays the role of the Hubble parameter at the beginning of the inflation phase. The moment $t_0>0$ is the conditional moment of the beginning of the inflation phase.

From solutions (\ref{Solution_H})-(\ref{Solution_p}) it is clear that case (i) corresponds to the value $\varkappa(t)=1$, case (ii) --- the value $\varkappa(t)=4/3$, and case (iii) --- the value $\varkappa(t)=0$.

We indicate some of the most obvious properties of the solution (\ref{Solution_H}), (\ref{Solution_e}),
(\ref{Solution_p}). The estimates given below are very rough. Let us accept the following estimates for the duration of the inflation time $t_{\mbox{inf}}$, and for the constant $\Lambda_0$:
\begin{gather}
t_{\mbox{inf}}\cong10^{-37}\mbox{s}, \quad H_0\cong10^{39}\mbox{s}^{-1}.
\label{Estimates}
\end{gather}
In (\ref{Estimates}) dimension is restored. Then $H_0t_{\mbox{inf}}\cong100$. 

Estimates of the Hubble parameter in the inflation phase differ greatly in different sources. We adopted the value $H\sim 1.7\cdot10^{15}\mbox{GeV}$, which is equivalent to the value (\ref{Estimates}) \cite{linde1996recent}. In the paper \cite{starobinsky2000future} the value of the Hubble parameter at the end of the inflation phase is estimated as $\sqrt{G}H<10^{-5}$. This value corresponds to $H\sim10^{38}\mbox{cm}^{-1}$, which is close to the value (\ref{Estimates}).

Let us take $\varkappa(t_0)\cong1/150$ and assume that during time $t_{\mbox{inf}}$ the function $\varkappa$ changes insignificantly. This assumption means (see the first of Eq. (\ref{Solution_H})) that in the inflation phase $|\dot{H}|\ll H^2$. The last inequality is a necessary condition for inflation \cite{motohashi2015inflation}. Then in the time interval $t_0<t<t_0+t_{\mbox{inf}}$ the solutions (\ref{Solution_H}), (\ref{Solution_e}), (\ref{Solution_p}) take the form
\begin{gather}
H(t)\cong H_0, \quad \tilde{\varepsilon}(t)\cong -\tilde{p}\cong-\Lambda_0+3H_0^2.
\label{Solution_t=0}
\end{gather}
Thus, during inflation, the scale factor $a(t)$ increased by  $(\exp H_0t_{\mbox{inf}})\approx(\exp100)\approx10^{43}$ times.

Assume that when $t>t_0+t_{\mbox{inf}}$, the function $\varkappa(t)$ becomes equal to $\varkappa=4/3$.
In this case, the solutions (\ref{Solution_H}), (\ref{Solution_e}),
(\ref{Solution_p}) give a power-law expansion:
\begin{gather}
H(t)\cong\frac{1}{2t}, \quad \tilde{\varepsilon}(t)\cong-\Lambda_0+\frac{3}{4t^2},
\quad \tilde{p}\cong\Lambda_0+\frac{1}{4t^2}.
\label{Power-law_expansion}
\end{gather}
Solution (\ref{Power-law_expansion}) shows that the scale factor and the density of real matter change according to the well-known law, as well as the correct equation of state in the case of ultrarelativistic matter (in (\ref{Real_matter}) dimension restored):
\begin{gather}
a(t)\propto\sqrt{t}, \quad \rho_{\mbox{real}}=\frac{3}{32\pi Gt^2}, \quad p_{\mbox{real}}=\frac13\varepsilon_{\mbox{real}}.
\label{Real_matter}
\end{gather}

From the equalities (\ref{Solution_e}) and (\ref{Solution_p}) it is clear that as we approach the point of the Big Bang, the quantities $|\tilde{\varepsilon}|$ and $\tilde{p}$ decrease and become equal to
\begin{gather}
\tilde{\varepsilon}=-(\Lambda_0-3H_0^2), \quad \tilde{p}=\Lambda_0-3(1-\varkappa(t_0))H_0^2
\quad \mbox{for}  \quad t=t_0.
\nonumber
\end{gather}
If we assume that at $t\longrightarrow0$ we have $\varkappa\longrightarrow0$ and $3H_0^2\longrightarrow\Lambda_0$, then the energy-momentum tensor tend to zero at the point of the Big Bang. However, it was shown above that such a continuation of the solution to the Big Bang point is impossible due to the development of quantum fluctuations.
Nevertheless, the tendency towards an absolute decrease in the energy-momentum tensor in the classical solution when approaching the Big Bang point shows the consistency of the presented classical and quantum approaches.

Let us estimate the temperature $T_c$ of the phase transition that violates PT symmetry. For this, we use the result of Volovik
\cite{volovik2009sitter,volovik2024sitter,volovik2024thermodynamics,volovik2023sommerfeld}, who calculated the local temperature of the "vacuum" in de Sitter space:
\begin{gather}
T_{\mbox{vac}}=\frac{\hbar H}{\pi }.
\label{Temperature_vac_deSit}
\end{gather}
Although in the theory under study the Hubble parameter, unlike in the case of de Sitter space, is not constant, here we will adopt the formula (\ref{Temperature_vac_deSit}) to estimate the temperature
(see also Section V).

It has been shown that at the phase transition point the classical Einstein equations (\ref{Einstein_eq_2}) are not valid. However, we use them only for a qualitative estimation. Since at the phase transition point the mean of the fermion energy-momentum tensor is zero ($\varepsilon=0$), then according to the second equation (\ref{Einstein_eq_2}) we have $H_c\sim\sqrt{\Lambda_0}$. From here and using (\ref{Temperature_vac_deSit}), (\ref{Estimation_1}) we obtain the estimate:
\begin{gather}
T_c\sim\frac{\hbar c}{l_P}\sim10^{18}\mbox{GeV}, \quad \mbox{or} \quad T_c\sim 10^{31}\mbox{K}.
\label{Temperature_c}
\end{gather}
The phase transition temperature can also be estimated as the energy of the Dirac sea enclosed in the Planck volume $V_P\sim l_P^3$: $T_c\sim(\hbar c/l_P^4)l_P^3\sim\hbar c/l_P$. This temperature is of the same order of magnitude as the temperature of the Grand Unification.

Let us also estimate the temperature in degrees Kelvin at the beginning of the inflation phase, when according to (\ref{Estimates}) $H_0\sim10^{39}\mbox{s}^{-1}$. Then
\begin{gather}
T_0\sim\frac{\hbar H_0}{k}\sim10^{28}\mbox{K}.
\label{Temperature_2}
\end{gather}

\subsection{A brief review of the divergent cosmological constant problem}

The problem with the cosmological constant is that the energy density of quantum zero-point fluctuations of the vacuum diverges as the fourth power of the cutoff parameter, and there is currently no generally accepted solution to compensate for this enormous energy density.

It seems to us that in this Section above a possible solution to the problem is presented, which is correct in the case when space-time has the property of granularity (lattice) on an ultra-small scale. Indeed, the introduction of a finite bare cosmological constant leads to a completely reasonable solution to Einstein’s equations: in the initial phase we have an exponential expansion of the Universe (inflation regime), which turns into a known expansion according to a power law in the regime of ultrarelativistic matter.

We consider it appropriate to give here a very brief and incomplete overview of attempts to solve the identified problem within the framework of traditional quantum field theory.

In the fundamental review \cite{weinberg1989cosmological} the following facts were stated regarding the problem of divergent energy (divergent cosmological constant) of the ground state in quantum field theory: (i) In flat Minkowski space-time, these divergences, generally speaking, take place, but in the case of supersymmetric theories, the energies of the ground states strictly vanish. (ii) In curved space-time, even in the case of supergravity, the cosmological constant diverges. (iii) The superstring theory does not save the situation either.

Of the later and specialized works, we note the works \cite{krotov2011infrared,polyakov2012infrared,akhmedov2014lecture,akhmedov2019characters,akhmedov2021curved,
kamenshchik2022massive}.
In these papers, efforts were made to solve the problem of the cosmological constant in detail, that is, through microscopic analysis. In particular, the probability of the following process was calculated. Let there be a massive particle in the de Sitter space. This particle gives rise to a pair of the same particles for a sufficiently long period of time. This problem has been studied for both free and interacting fields. A similar statement of the problem for massive charged particles in the case of a flat space-time in the presence of a constant electric field leads to the creation of particle-antiparticle pair that weaken the initial electric field. In the case of de Sitter space, pair production also leads to a decrease in the cosmological constant with time. Unfortunately, in these works there is no study of the reverse influence of quantized material fields on the space-time geometry. It is possible that continued efforts in this direction will lead to a solution to the problem of the cosmological constant.

In the paper \cite{zel1972particle} the mean  of the energy-momentum tensor of a quantized scalar field is calculated in the case of an anisotropic metric, which is considered to be classical and variable in time.
Regularization is carried out in the usual way: the vacuum expectation value of the energy-momentum tensor, calculated in the case of a stationary vacuum, is subtracted from the obtained value.

The authors of the paper \cite{kamenshchik2018pauli} study such models of field theory which, although not supersymmetric, have the same number of boson and fermion degrees of freedom. In this case, the divergences of the highest, fourth degree are eliminated in the quantum mean of the energy-momentum tensor.
It is shown what conditions the already renormalized field masses must satisfy in order to reduce all other divergences. This approach can be called a method of fine-tuning the theory, as a result of which the infinite vacuum energy disappears.

The work \cite{appleby2020well} seems to us to be interesting and complementary to the present work, since a bare cosmological constant is also introduced in \cite{appleby2020well}, and the reduction of the huge vacuum energy is a dynamic effect, not a fine-tuning effect.

Another interesting approach to solving the problem, using the macroscopic thermodynamic ideology, is presented in \cite{klinkhamer2022big} (see there also references for the articles of F.R. Klinkhamer and G.E. Volovik). The main idea of this approach is as follows.  If the system comes to a state of thermodynamic equilibrium, then a large thermodynamic potential is of interest. Let $\Omega$ be a large thermodynamic potential for the spatial volume  $V$.
It is known that
\begin{gather}
\Omega(\beta,\mu,V)=-P(T,\mu)V.
\label{Large_therm_pot}
\end{gather}
In the case of thermal equilibrium (if it exists), the pressure on the right side of the equality (\ref{Large_therm_pot}) tends to zero, since there is no external pressure at all. Further, the effective energy-momentum tensor of matter is formed by the potential
(\ref{Large_therm_pot}). Therefore, the effective energy density of matter, including the vacuum energy, under the condition of thermal equilibrium is estimated as $\varepsilon\sim\Omega/V\longrightarrow0$.
Thus the problem of the divergent cosmological constant is removed.

In the work \cite{wang2017huge} the use of the metric (\ref{Metric}) is criticized due to the strong fluctuations of all fields on the Planck scales of frequencies and wavelengths.
This conclusion is made on the basis of studying the correlators of the energy-momentum tensor components constructed from quantized matter fields. The calculation shows that even in the case of free quantized fields, the vacuum average values of the energy-momentum tensor components and their fluctuations are of the same order and diverge as the fourth power of the cutoff momentum.
The fact that the divergences of these quantities are of the fourth order means that these divergences saturate on the scale of the cutoff momentum. In other words, exclusively high-frequency degrees of freedom contribute to the discussed averages.
Therefore, we draw the following conclusion: if we describe the dynamics of any degrees of freedom that have frequencies and wave vectors much less than Planck's, then the high-frequency fluctuations discussed in \cite{wang2017huge} do not play a role in such dynamics. Indeed, the correlators between the low-frequency and high-frequency degrees of freedom are in fact equal to zero. This justifies the use of the metric (\ref{Metric}) in the Einstein equations  discussed above, as well as the assumption that the energy-momentum tensor is uniform in space.
However, these assumptions become incorrect when the scale factor and mean of the energy-momentum tensor components become  small and the temperature becomes high enough (see Section III).

\section{Thermodynamic considerations}

The purpose of this Section is to qualitatively understand the result of temperature calculation (\ref{Temperature_vac_deSit}). It will be shown that the local vacuum temperature in a dynamic Universe with a changing scale factor is always of the order of $\hbar H$.

The estimation (\ref{Estimates}) means that
\begin{gather}
H_0^2\ll\Lambda_0.
\label{Thermodyn_Equilibrium_condition}
\end{gather}
It can be seen from Eq. (\ref{Solution_e}) that the maximum frequencies of the degrees of freedom of matter in the modern era are of the order of
\begin{gather}
|\omega_{\mbox{max}}|\sim \sqrt{\Lambda_0}.
\label{Frequencies_max}
\end{gather}
Consider the time interval $\Delta t\lesssim H^{-1}$, for which we have
\begin{gather}
\Delta a/a\sim H\Delta t\lesssim1, \quad \Delta t|\omega_{\mbox{max}}|\gg1.
\label{Time}
\end{gather}
Taking into account Eq. (\ref{Time}), we can assume that for a time interval $\Delta t$ the thermodynamic equilibrium of the vacuum degrees of freedom is realized.
This assumption cannot be extended to those degrees of freedom whose frequencies $\Delta t|\omega|\lesssim1$.
But such degrees of freedom make a small contribution to the total energy-momentum tensor. Thus, the degrees of freedom with frequencies in the interval
\begin{gather}
H_0\sim\Delta t^{-1}\ll|\omega|<|\omega_{\mbox{max}}|
\label{Frequencies_interval}
\end{gather}
are in a state of thermal equilibrium or close to it.

The conclusion about the thermalization of vacuum degrees of freedom in the indicated frequency range (\ref{Frequencies_interval}) can be supported by the following reasoning.
There is no indication that the system under study is integrable. Therefore, the dynamics of all distribution functions of the system must be described by the Boltzmann equation. The Boltzmann equation brings the system to a locally equilibrium state in time intervals commensurate with the inverse values of the characteristic frequencies of the dynamic degrees of freedom. This statement is true under the condition of an insignificant change in external conditions during the process of establishing equilibrium.
To fulfill the last condition, it is necessary to satisfy the estimates (\ref{Time}) and (\ref{Frequencies_interval}).

When passing to the Euclidean signature by Wick's rotation $\Delta t=-i\Delta\tau$ , the parameter
\begin{gather}
 T_{\mbox{vac}}\equiv\beta^{-1}=\hbar(\Delta\tau)^{-1}\approx\hbar H
\label{Temperature_1}
\end{gather}
acquires the meaning of {\it local} vacuum temperature. Here the dimension is restored. An exact calculation of the local vacuum temperature in de Sitter spacetime has been done in \cite{volovik2009sitter,volovik2024sitter,volovik2024thermodynamics,volovik2023sommerfeld}: $T_{\mbox{vac}}=\hbar H/\pi$. From the method of calculating the temperature in the cited works it is evident that the calculation result is valid when $|\dot{H}|\ll H^2$. The last restriction takes place in the inflation phase. Local temperatures differ from the horizon temperature in de Sitter space-time \cite{gibbons1977cosmological}, although all the temperatures discussed here are of the same order. As can be seen, simple qualitative reasoning within the framework of quantum field theory leads to a correct estimate of the local vacuum temperature.

Let's make a remark. The calculation of the local vacuum temperature in the paper   \cite{volovik2009sitter} is tied to the space-time of a particular geometry: the de Sitter space-time. On the contrary, our approach to estimating the local vacuum temperature can be applied to the space-time of any geometry. We need only relatively local homogeneity and isotropy of space. The general conditions for the existence and measurability of the local vacuum temperature are not discussed here.

Here one should cite the classic work of Unruh \cite{unruh1976notes}, in which the problem of calculating the temperature of, for example, the horizon of a black hole, was connected with the dynamics of quantized fields.

Once again, we note that thermodynamic considerations do not apply to low-frequency degrees of freedom. In particular, ordinary real matter may, generally speaking, not be in a state of thermal equilibrium.

It can be seen from the first Eq. (\ref{Einstein_eq_1}) that the constant $\Lambda_0$ cancels out the huge negative energy of the vacuum, so that in the era of power-law expansion only the relatively extremely small positive energy density of real matter affects the dynamics.  From the given solution of Einstein's equations, it can be seen that the huge value of pressure is also mainly reduced by the constant $\Lambda_0$.  In the presented solution we have $\tilde{p}\sim-\tilde{\varepsilon}\sim\Lambda_0$. Such a ratio of energy and pressure is possible only in the general theory of relativity.

Let us introduce the notation $\tilde{\epsilon}_{\mbox{vac}}\equiv(\tilde{\varepsilon}+\Lambda_0)$.
With the help of Eqs. (\ref{Rescaling}) and (\ref{Einstein_eq_2}) we find:
\begin{gather}
\tilde{\epsilon}_{\mbox{vac}}\equiv(\tilde{\varepsilon}+\Lambda_0), \quad
\epsilon_{\mbox{vac}}=\frac{\tilde{\epsilon}_{\mbox{vac}}}{8\pi G}=\frac{3H^2}{8\pi G}.
\label{Notation_vac}
\end{gather}
The last relation is contained in \cite{volovik2023sommerfeld}. It is the quantity (\ref{Notation_vac}) that interacts with the Einstein tensor in the Einstein equation. According to (\ref{Temperature_1}) the vacuum temperature $T\approx\hbar H$. Therefore, the ratio (\ref{Notation_vac}) is rewritten as follows:
\begin{gather}
\epsilon_{\mbox{vac}}=\lambda\frac{3T^2}{8\pi G\hbar^2}, \quad \lambda\sim1.
\label{Energy_Dens}
\end{gather}
\\

{\it Statement 2}. If we know the energy density as a function of temperature, then the free energy density is
\begin{gather}
f_{\mbox{vac}}= -T\int_0^T\frac{\epsilon_{\mbox{vac}}(T')}{(T')^2}\d T'.
\label{Free_Energy_Dens}
\end{gather}
\\

Checking the Statement 2 is reduced to checking the fact that the function (\ref{Free_Energy_Dens}) satisfies the following system of equations:
\begin{gather}
s_{\mbox{vac}}=-(\partial/\partial T)f_{\mbox{vac}}, \quad
f_{\mbox{vac}}=\epsilon_{\mbox{vac}}-Ts_{\mbox{vac}}.
\label{Free_Energy_Def}
\end{gather}
Here $s_{\mbox{vac}}$ is the local vacuum entropy density.
In our case (\ref{Energy_Dens}) we obtain:
\begin{gather}
f_{\mbox{vac}}=-\epsilon_{\mbox{vac}}=-\lambda\frac{3T^2}{8\pi G\hbar^2},  \quad
s_{\mbox{vac}}=\lambda\frac{3T}{4\pi G\hbar^2}.
\label{Our_case}
\end{gather}

A more accurate calculation for the quantities (\ref{Energy_Dens}) and (\ref{Our_case}) in de Sitter space-time is given in \cite{volovik2023sommerfeld}, which shows that $\lambda=\pi^2$.

 The estimate $\tilde{p}\cong\Lambda_0$ together with the vacuum energy hypothesis $\tilde{\varepsilon}\cong-\Lambda_0$ justifies the equation of state (\ref{State_equation}).
In both parts of equality $(\tilde{\varepsilon}+\tilde{p})=\varkappa(\tilde{\varepsilon}+\Lambda_0)$, the diverging values of the quantities cancel each other out. This fact is the result of solving dynamic equations.

According to Eqs. (\ref{Solution_H}) and (\ref{Temperature_1}) we have:
\begin{gather}
\hbar\d\beta\sim\d(1/H)=3/2\varkappa\d t.
\nonumber
\end{gather}
But in the inflation phase $a(t)=a_0e^{Ht}$, and so $\d t=H^{-1}\d a/a$. Thus we have:
\begin{gather}
\d\beta/\beta\sim(3/2)\varkappa\d a/a.
\label{Temperature_change}
\end{gather}
Since the temperature decreases in the inflation phase, it can be seen from (\ref{Temperature_change}) that
$\varkappa(t_0)>0$.

\section{Conclussion}

In the works \cite{diakonov2011towards,vladimirov2012phase,vladimirov2014diffeomorphism,volovik2021dimensionless} and \cite{vergeles2021note} the equality
\begin{gather}
\left\langle e_{{\cV}_1{\cV}_2}^a\right\rangle=\varkappa^{(0)}_{({\cV}_1{\cV}_2)}\left\langle \Theta^a_{{\cV}_1{\cV}_2}\right\rangle
\label{Lat_tetrad_Aver}
\end{gather}
was proposed. This equality holds both in the lattice theory of gravity, identical to the one studied here, and in the continuous theory.  The relation (\ref{Lat_tetrad_Aver}) is justified by the fact that the values under the average transform  identically under all symmetry transformations, including discrete PT symmetry. The relation (\ref{Theta_Mean_Minkov}), obtained by direct calculation in the continuum limit at zero temperature, is another argument in favor of the validity of the relation (\ref{Lat_tetrad_Aver}).

Thus, if the hypothesis proposed above is correct, then at the maximum possible temperature, the right side of Eq. (\ref{Lat_tetrad_Aver}) vanishes according to (\ref{Mean_Zero}), whence follows
\begin{gather}
\left\langle e_{{\cV}_1{\cV}_2}^a\right\rangle_{\mbox{Gauge Fix}}=0.
\label{Lat_tetrad_Zero}
\end{gather}
This result is obtained directly in the same way as the result (\ref{Mean_Zero}) was obtained. The last equality means that in the lattice theory of gravity in the high-temperature phase, space collapses into a point.

However, in the long-wave limit at low temperature, the energy-momentum tensor of the Dirac field is non-zero (\ref{Theta_Mean_Minkov}), the quantized field of the tetrad fluctuates weakly and its mean is not equal to zero.

The above means that in the lattice theory of gravity under study there is a phase transition in temperature (possibly more than one). In the high-temperature phase, space is curled up into a point, the average of the energy-momentum tensor is zero, and the discrete PT symmetry is not broken. On the contrary, in the low-temperature phase these quantities are not equal to zero and PT symmetry is broken. The role of the order parameter is played by the average $\langle e^a_{\mu}\rangle$, which becomes non-zero in the low-temperature phase. In the low-temperature phase, the process of exponential expansion of space begins, transitioning to a power law of expansion. 
During the phase transition from the high-temperature to the low-temperature phase, domains with opposite mean values $\langle e^a_{\mu}\rangle$ and $\langle \Theta^a_{\mu}\rangle$ can be formed. The domain wall separating such domains was studied in \cite{vergeles2021domain}

\begin{acknowledgments}

I thank Prof. G.E. Volovik for awakening my interest in the thermodynamic study of the problem.
I am grateful to Prof. E.T. Akhmedov for numerous discussions and advice in the course of work.
This work was carried out as a part of the State Program 0033-2019-0005.

\end{acknowledgments}


\end{document}